\documentclass[acmsmall]{acmart}

\AtBeginDocument{%
  \providecommand\BibTeX{{%
    \normalfont B\kern-0.5em{\scshape i\kern-0.25em b}\kern-0.8em\TeX}}}

\usepackage{algorithm}
\usepackage{algorithmic}
\usepackage{amsmath}
\usepackage{amsfonts}
\usepackage{float}
\usepackage{graphicx}
\usepackage[utf8]{inputenc}
\usepackage{mathtools}
\usepackage{microtype}
\usepackage{multirow}
\usepackage{stmaryrd}
\usepackage{tcolorbox}
\usepackage{textcomp}
\usepackage{xcolor}

\DeclareMathOperator*{\argmax}{argmax}

\copyrightyear{2022}
\acmYear{2022}
\setcopyright{rightsretained}

\acmJournal{PACMHCI}
\acmYear{2022} \acmVolume{6} \acmNumber{MHCI} \acmArticle{202} \acmMonth{9} \acmPrice{}\acmDOI{10.1145/3546737}

\usepackage{etoolbox}
\makeatletter
\patchcmd{\maketitle}{\@copyrightpermission}{
   \begin{minipage}{0.2\columnwidth}
     \href{https://creativecommons.org/licenses/by/4.0/}{\includegraphics[width=0.90\textwidth]{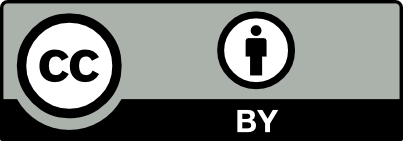}}
   \end{minipage}\hfill
   \begin{minipage}{0.8\columnwidth}
     \href{https://creativecommons.org/licenses/by/4.0/}{This work is licensed under a Creative Commons Attribution International 4.0 License.}
   \end{minipage}

   \vspace{5pt}
}{}{}

\makeatother

\acmSubmissionID{V6mhci202}

\begin{document}

\title{Spatial model personalization in Gboard}

\author{Gary Sivek}
\affiliation{%
  \institution{Google LLC}
  \city{New York}
  \state{NY}
  \country{USA}}
\email{gsivek@google.com}

\author{Michael Riley}
\affiliation{%
  \institution{Google LLC}
  \city{New York}
  \state{NY}
  \country{USA}}
\email{riley@google.com}

\renewcommand{\shortauthors}{Gary Sivek \& Michael Riley}



\begin{abstract}
We introduce a framework for adapting a virtual keyboard to individual user behavior by modifying a Gaussian spatial model to use personalized key center offset means and, optionally, learned covariances. Through numerous real-world studies, we determine the importance of training data quantity and weights, as well as the number of clusters into which to group keys to avoid overfitting. While past research has shown potential of this technique using artificially-simple virtual keyboards and games or fixed typing prompts, we demonstrate effectiveness using the highly-tuned Gboard app with a representative set of users and their real typing behaviors. Across a variety of top languages, we achieve small-but-significant improvements in both typing speed and decoder accuracy.
\end{abstract}

\maketitle

\section{Introduction}

Gboard is a widely-used keyboard app for Android and iPhone. In \citep{ouyang2017mobile} the authors describe the basic functionality, a few years out of date but still largely relevant: when a user taps keys to spell out an input word, the Gboard decoder converts them to bikeys, generates candidates, and scores, prunes, and ranks them through a combination of a language model score and a spatial model score. Newer work since then includes modules that control auto-correction decisions through more than just raw language and spatial model scores, and a key correction model to decide when a key press near a boundary should be forcibly corrected to a key on the other side of that boundary. None of the details of the language model, auto-correction module, or key correction module are germane here beyond that they exist and all improve typing accuracy.

The spatial model is based on a fixed Gaussian model with distances computed from physical centers of keys; we say "based on" because there are other non-Gaussian costs introduced to support edits, word completions, etc. This work aims to personalize the score to account for most users' tendency to touch keys with non-zero offsets that vary by key.

\subsection{Related work}

Prior studies have evaluated assumptions about typing habits and systemic improvements to decoding in a number of directions:

{\bf Touchpoint distribution.} Azenkot and Zhai in \citep{azenkot2012touch} used transcription tasks on a specially-designed Keyboard Touch Collector app to study touch behavior as a function of posture, showing clear evidence of bivariate Gaussian distributions per user and key. Zhu et al in \citep{zhu2018typing} took this further, training per-key Gaussian statistics on an {\it invisible} keyboard to show that such a keyboard was feasible as an input method; they saw WPM (words per minute) increasing from 26.29 to 29.30 with an adapted invisible keyboard instead of an unadapted one.

{\bf Posture.} Dhakal et al (\citep{dhakal2018observations}) and Palin et al (\citep{palin2019people}) collected large-scale typing data sets from distributed transcription tasks and analyzed typing performance as a function of typing speed, {\it posture} (i.e., hands and fingers used), and other factors. While the former ran primarily on physical keyboards, the latter used a self-selected set of more than 37,000 online volunteers performing web-based transcription tasks in English with their own mobile device and virtual keyboard. They found that posture affects typing speed: typing with two thumbs (38.0 WPM) is faster than typing with two index fingers (32.6 WPM), which is faster than typing with just one finger of any type (25.0-30.8 WPM). They do not, however, break down uncorrected error rate by posture. Similarly, Reyal et al (\citep{reyal2015performance}) in a series of experiments with 12 volunteers found higher speed and lower error rates for touch typing with two thumbs than with other postures.

{\bf Personalization.} Henze et al (\citep{henze2012observational}) published a typing game to learn touch point distributions, collecting nearly 48 million keystrokes from 73,000 installations, then found that a per-device-type function shifting touch points toward key centers improved speed by 2.6\% and error rate by 7.7\% over a control. Weir et al in \citep{weir2012user} used Gaussian Process regression and a touchscreen game to learn intended touch points from actual sensor data, showing significant accuracy gains for 8 participants typing 1000 keystrokes apiece with fixed orientation and posture. Baldwin and Chai in \citep{baldwin2012towards} recruited 8 volunteers to perform long chats in pairs on a custom keyboard, then simulated typing with key-target resizing offline. They found that restricting training data for personalization to in-vocabulary words gave the best performance, reducing keystroke error rate from 4.8\% to 4.3\%.

{\bf Posture-based personalization.} The WalkType system (\citep{goel2012walktype}) found that incorporating accelerometer data into a model to account for motion-based impairment improved typing speed from 28.3 to 31.3 WPM and decreased uncorrected errors from 10.5\% to 5.8\% when compared to a model that simply chooses the key whose boundaries contain a given touchpoint, also finding that this model was even useful when a user was sitting. The ContextType system (\citep{goel2013contexttype}) inferred posture using tap sizes and time between taps, and used that to build per-posture personalized spatial models; they found a 20.6\% reduction in error rates when compared to personalized spatial models that do not account for posture, but no change in typing speed due to an increase in {\it corrected errors}, which took time to correct. Both studies used a set of 16 participants typing 30 phrases each. Musi{\'c} and Murray-Smith (\citep{music2016nomadic}) later confirmed, via 20 volunteers playing a mobile game while walking, that touch point accuracy degrades while walking, but that posture-dependent offset models outperformed static offset models in error reduction (14-30\% vs 6-17\%).

Key takeaways from all this past research are that spatial models which depend on user and posture can improve performance, and that per-key typing behavior tends to follow Gaussian distributions. Most training data comes from a combination of artificial games or transcription tasks, and keyboards tested tend to be either ancient (in smartphone terms), simplified (especially in comparison to modern soft keyboards), or invented strictly for the sake of the research. Notably, Reyal et al in \citep{reyal2015performance} do demonstrate a difference between typing performance in the lab versus "in the wild," but the "wild" setup still involved transcription tasks interspersed between normal daily tasks. In recognition of this, Buschek et al (\citep{buschek2018researchime}) built ResearchIME, which occasionally asks the user to report posture but otherwise simply logs typing events with frequent redactions for privacy. They then collected nearly 6 million keyboard events, including nearly 1 million keystrokes, from 30 volunteers typing normally over the course of three weeks to report real-world usage metrics. But even this is a small, non-representative sample of users, and they do not use this data to personalize the keyboard.

In contrast to past work, the work presented here represents, to our knowledge, the first spatial personalization results on a production-quality soft keyboard trained with and evaluated on realistic data.

\section{Background}

Gboard has a number of input modes: next-word prediction, suggested word completions, and of course direct input of a word. The two main forms of direct input are {\it tap} and {\it gesture}. Tap mode means touching the screen once per input character, allowing additional taps for layout switches, long presses for variants such as diacritics, and edits. Gesture mode involves one continuous glide along the screen to type the entire word; see, e.g., \citep{zhai2012word}. This paper is concerned only with tap typing.

As a user types a word, we update a search to find the best candidate(s) for the completed word. Given a left context $w_1 \cdots w_n$ and current-word touch points $\vec{z}_1 \ldots, \vec{z_k}$, the best candidate word $\hat{w}$, analogous to the noisy channel model of speech recognition (\citep{jelinek1998statistical}) but with the acoustic inputs replaced by geometric touch inputs, is:

\begin{align}
\label{eqn:fundamental}
\hat{w} & = \argmax_w \frac{p(\{\vec{z_i}\} | w)\, 
p(w | \{w_j\})} {p(\{\vec{z_i}\})}\\
& = \argmax_w \log\nonumber p(\{\vec{z_i}\} | w) + \log p(w | \{w_j\})\\
& = \argmax_w SM(w; \{\vec{z_i}\}) + LM(w; \{w_j\})\nonumber
\end{align}

where $SM$ is the spatial model score and $LM$ is the language model score. The language model is described in more detail in \citep{ouyang2017mobile}. The spatial model score is a log probability, so we may write it as a sum over the touch point sequence of corresponding Gaussian exponent terms. If a key press is determined to be an insertion, deletion, or transposition, we add a pre-determined fixed cost. Otherwise, there is a pre-determined fixed {\it substitution cost} and a Gaussian standard deviation term $\sigma$. Let the touch point be $\vec{z_i} = (x_i, y_i)$ and let the corresponding character of the word have key center $(\hat{x}_i, \hat{y}_i)$ and dimensions $(w_i, h_i)$. We write the spatial model summand for this touch point as:

\begin{equation}
\label{eqn:gaussian}
min\left\{\frac{\Delta x_i^2 + \Delta y_i^2}{2\sigma^2}, \textrm{substitution cost}\right\}
\end{equation}

where

\begin{equation}
\label{eqn:offset}
\Delta x_i = \frac{x_i - \hat{x}_i}{w_i},\ \Delta y_i = \frac{y_i - \hat{y}_i}{h_i}.
\end{equation}

Both the fixed costs and $\sigma$ were determined empirically at one point in time to minimize word error rate on a collection of test sets.

The key observation here is that the key centers $\{(\hat{x}_i, \hat{y}_i)\}$ were chosen to model expected typing behavior, but do not match in practice. We ran user studies with predetermined typing prompts and saw a variety of typing patterns.

\begin{figure}[H]
    \includegraphics[width=0.5\linewidth]{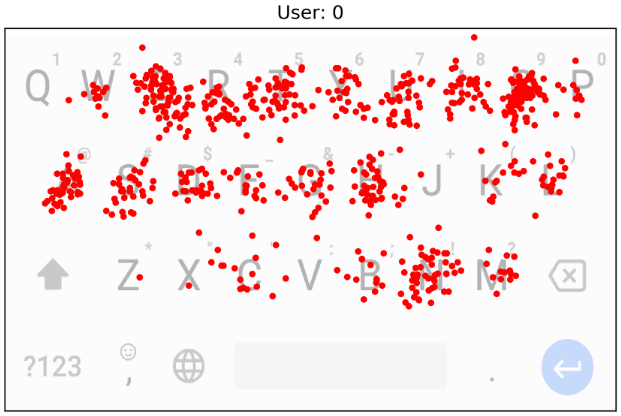}
    \caption{Touch points of user study user \#0.}
    \label{fig:user0}
    \Description{User 0 touch points. Fully described in the text.}
\end{figure}

\begin{figure}[H]
    \includegraphics[width=0.5\linewidth]{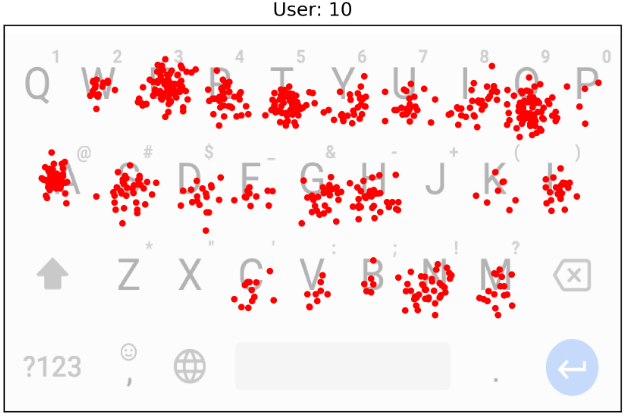}
    \caption{Touch points of user study user \#10.}
    \label{fig:user10}
    \Description{User 10 touch points. Fully described in the text.}
\end{figure}

\begin{figure}[H]
    \includegraphics[width=0.5\linewidth]{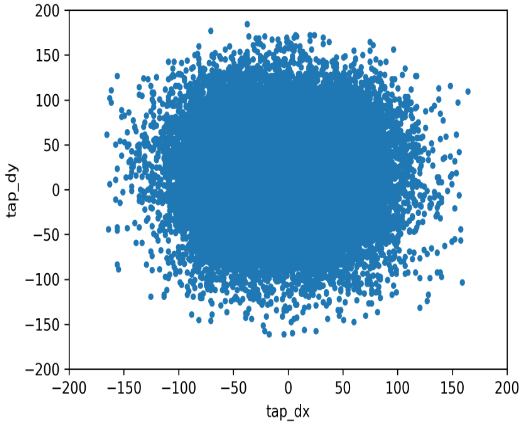}
    \caption{All touch offsets from all users.}
    \label{fig:alltouches}
    \Description{Global scatter plot of key touch point offsets. Fully described in the text.}
\end{figure}

User 0 in Figure \ref{fig:user0} and user 10 in Figure \ref{fig:user10} show very distinct patterns, but they importantly rarely center their touches for {\it any} key in the physical key center. This suggests that we could improve the key probability estimates by computing distances from better points than the physical centers. Differences between users suggest that they should be personalized; see, for example, how user 0 tends to miss [T] to the left while user 10 misses down and a little to the right. Also, note that users miss different keys in all different directions; the plot of all touch offsets across all users is shown in Figure \ref{fig:alltouches}, and it's roughly centered at 0. Consequently, a single global Gaussian has not only the wrong center but the wrong variance. We should reconsider our choice of $\sigma$ as we push key centers in the right direction.

\section{Experiment setup}\label{sec:setup}

See \citep{tang2010overlapping} for a discussion of experiment infrastructure and metrics; we follow that, but include more specific details here.

\subsection{Training data}

We use an on-device adaptation framework within Gboard which keeps data on the phone's persistent storage; none of the touch data described here or typed words ever leave the device. We also want to adapt to changes in typing habits and to limit device disk usage, so we have older data expire by keeping at most the $N$ most recent touch points ($250 \leq N \leq 800$ in all experiments; see the Results section for details). Another initial consideration was that smaller histories might adapt to changes in typing posture and/or orientation more readily, but it seems unlikely that such temporary changes would be handled by any history large enough to be useful. Although we have access to orientation information, we did not use it, and we did not attempt to predict posture at any point.

For privacy reasons, we store touch data in coarse {\it buckets}, where a ``bucket" means a collection of touches which are contiguous. All touches in an older bucket necessarily precede touches in a newer bucket, but touches within a bucket are unordered rather than in exact sequence so the typed content cannot be reconstructed even with physical access to the device. When old data expires, the entire oldest bucket will then expire at once.

Per-user training data consists of all touch points from committed words such that the word (a) was committed via tap typing, and (b) matches the final committed text for that text span. We collect touch point statistics in a map:
$${\tt Stats(x,y)} = \left\{ \sum 1, \sum \Delta x, \sum \Delta y \right\}$$

where $(x, y)$ is the physical center of a particular key, and the sums span all characters typed in non-expired training data using that key. (In practice, we keep
complete Gaussian sufficient statistics by including $\sum (\Delta x)^2, \sum (\Delta y)^2$, and $\sum (\Delta x)(\Delta y)$ for use in computing covariance, but that's not relevant for now.) 

\begin{figure}[H]
    \includegraphics[width=0.5\linewidth]{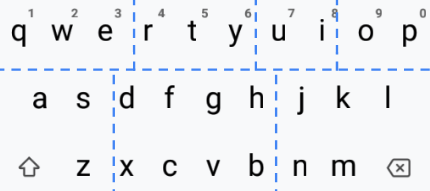}
    \caption{A sample clustering of keys.}
    \label{fig:cluster}
    \Description{A keyboard layout showing a possible clustering into seven groups. The top row is split into Q-W-E, R-T-Y, U-I, and O-P clusters; the bottom left cluster is A-S-Z; the bottom middle is D-F-G-H-X-C-V-B; and the bottom right cluster has J-K-L-N-M.}
\end{figure}

To address data sparsity, especially for the less frequently typed keys, we partitioned the keyboard into a number of clusters. This was accomplished by periodically applying a greedy decision tree-based algorithm (see Algorithm \ref{Algorithm} in Appendix \ref{app:algorithm}) that finds key clusters along with personalized per-cluster key center offsets. Each cluster is a rectangular grouping of adjacent keys defined by a series of vertical and/or horizontal split lines. See Figure \ref{fig:cluster} for an example when 7 clusters were chosen. For a given clustering, we pool our collected statistics per cluster. This is then used to calculate personalized offsets and adjust the physical key center coordinates in equation \ref{eqn:offset}.

Training data collection on device was enabled for all users in advance of the studies in the next section, with the exception of those studies evaluating training data quantity. This enabled us to collect metrics immediately without a "warm-up" period.

\subsection{User selection}

The pool of eligible users for these studies consists of all Gboard users worldwide with sufficiently-updated APKs on Android. When results are specified per language, we take the appropriate geographically-restricted subset: for example, experiments with pt-BR are limited to those users in Brazil. The en-US results are the exception, as we split that into US-internal and US-external populations.

Each study consists of a subset of the eligible population, selected uniformly at random, which is subsequently apportioned uniformly at random into a large control population and one or more experiment arms, all running concurrently. The app itself gives no indication whether a user is in control or experiment.

All studies have a large control population using default behavior, and the only differences between the control and an experiment arm are the population size and the specified adaptation behavior. Both control and experiment run for identical periods of time, which is a whole number of contiguous days, always at least a week, but with different studies choosing different durations based on external factors like release schedules and competing studies. Users are apportioned between them in pre-determined proportions randomly and blindly. No two studies here ran on the same users at the same time, except that two concurrent studies might share a control population.

Not all studies, or even arms within a study, used the same population size and/or duration. This may be due to the number of simultaneous arms, the number of simultaneous studies diverting users away, or even the perceived importance of measuring results on a particular arm. It is additionally hard to measure user numbers precisely, so we will instead report total number of words committed as a reflection of the total size. No experiment arm here had fewer than 90 million words typed even after targeting a specific country and filtering by language, but most had far, far more.

\subsection{Metrics}

The metrics described in this section are computed for all typing by all included users over the duration of the study without regard to posture, device orientation, app in which the user was typing, or other factors, and thus fully reflect typical real-world usage. However, language-specific metrics provided here are restricted to words typed when Gboard reports that the active language matches. For example, all users in the United States are eligible for inclusion in the es-US experiments, but only words typed with the Spanish language model active are included in the reported metrics; this avoids dilution of the results. We do this by logging enough word event metadata like event type (word committed, auto-correction, edit, etc), active language, spatial model score, and timing info to compute and then aggregate these, but we do not store anything about the typed content itself or the touch points.

"Statistically significant" here will mean significant at a 0.05 level (null hypothesis: the relative change in metric is 0). There are three key metrics we will watch across all live experiments:

\begin{itemize}
    \item \textbf{Average spatial cost}. The average spatial {\it score} is the arithmetic mean of spatial model scores over all final committed words in log space; the average spatial {\it cost} is the additive inverse, which is nonnegative. It lacks a few pieces of context, in particular the $p(\{\vec{z_i}\})$ term in the denominator of the equation \ref{eqn:fundamental} and the magnitude of all non-substitution edit costs. Thus we cannot in general read much into the specific value or even relative changes. But in the special case where all non-Gaussian costs are fixed and the isotropic scale $\sigma$ is unchanged, a better-fitting set of per-key-cluster means should decrease the Gaussian summand while all other terms remain fixed, thus decreasing the spatial cost. In that case, all we will care about is which direction it moves and whether the move is statistically significant.

    \item \textbf{WMR}, or {\it Words Modified Ratio}. There are no exact labels, since this is not a transcription task and we never see the text users type, so we instead use this as a proxy for the traditional {\it word error rate}. WMR indicates what proportion of words were in some way changed after the initial commit, perhaps due to typos or incorrect auto-corrections which were reverted. It does not include mistakes that the user missed or decided not to fix, but it does include users deciding {\it post hoc} that an altogether different word would be better.

    \item \textbf{WPM}, or {\it Words Per Minute}. This is our best estimate of average user typing speed.
\end{itemize}

The primary mechanism for changing WMR and WPM is through auto-correction. We define the explicitly-tapped sequence of characters to be the {\it literal} candidate; if it is not part of the language model's word-based lexicon, a simple character-based backup language model attempts to assign it a somewhat reasonable score. The decision to auto-correct is then a function whose inputs include the literal's language model and spatial model scores and the top non-literal candidate's language model and spatial scores. For most languages at the time of the live studies, the decision was simply equivalent to whether $LM_{cand}+SM_{cand}$ was greater than $LM_{literal} + SM_{literal}$, in addition to some exogenous disqualifying criteria. Spatial model personalization shifted scores in both directions, eliminating some bad auto-corrections and inducing some additional good ones.

We also aim to answer several questions optimally:

\begin{itemize}
    \item How much training data do we need for offsets to be useful?
    \item How should we weight training data as it ages?
    \item Into how many clusters should we divide the keyboard?
\end{itemize}

\section{Results}

The studies described below ran serially due to the parameter space being too large for a comprehensive search. We evaluated the size, structure, and weighting of the training data, then fixed a single choice for subsequent experiments. Next, we evaluated the effect of the number of key clusters, found it made no significant difference, and fixed a single choice. Last, we explored a range of different isotropic Gaussian scale parameters per language. The product set of all these parameter values would have had our populations divided 210 ways per language, likely far too small for confidence in the resulting metrics. Future work might involve a set of experiments on a smaller local parameter grid to see if these values are, in fact, optimal.

\subsection{Training data}

The first and most fundamental question asked was how much training data do we need to learn typing habits. We also hoped to adapt quickly to changes in typing behavior, so we needed to know whether to down-weight older training data. Here, as in all studies, training data includes prior touch points regardless of orientation, posture, application, or other factors, and learned Gaussian statistics are similarly applied agnostically and across changes in these input factors.

\begin{figure}[H]
    \includegraphics[width=0.5\linewidth]{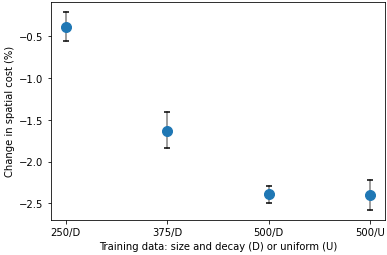}
    \caption{Training data size / decay, round 1.}
    \label{fig:decay}
    \Description{Effect of training data size on spatial cost, study #1. More training data means lower cost, but there is no difference between down-weighting older data and using uniform weights.}
\end{figure}

The first study tested different amounts of training data, using arithmetic decay to down-weight older buckets. Specifically, given statistics maps $\mathtt{Stats}_1, \ldots, \mathtt{Stats}_N$ for a collection of $N$ buckets $B_1, \ldots, B_N$ from newest to oldest, and an arithmetic decay rate $d$, we set:
$$w_i := \max(1 - (i-1)d, 0)$$
$$\mathtt{Stats}(x,y) =
 \left\{ \sum_{i=1}^N \left(w_i \sum_{B_i} \sum_{\mathrm{kc}=(x,y)} 1\right) \right. ,
 \sum_{i=1}^N \left(w_i \sum_{B_i} \sum_{\mathrm{kc}=(x,y)} \Delta x\right) ,
 \left. \sum_{i=1}^N \left(w_i \sum_{B_i} \sum_{\mathrm{kc}=(x,y)} \Delta y\right) \right\}$$

Each user's device stored the training data in $N=5$ equally-sized buckets with a decay rate of $d=0.15$ (excepting "500/U"), then used the personalized weighted touch offset data to compute 7 key clusters and their optimal offsets as described earlier, regularly updating using new data as the study continued. There was no change in isotropic Gaussian scale, so the expectation was that a better fit for Gaussian means would reduce the per-character Gaussian summand of the spatial cost.

This ran for 11 days and did not use geotargeting. The control population and "500/D" population typed 19B and 7.7B words, respectively, and all other arms typed 3.9B words each. The effect on spatial cost, shown in Figure \ref{fig:decay} as a relative change from average spatial cost in a large control population with no adaptation, is that more data leads to more accuracy. (The bars surrounding each point represent the 95\% confidence interval.) In addition to a 500-touch arm with arithmetic decay, we tested a 500-touch arm with uniform weights and found them to perform indistinguishably. Our choice of their experiment arms and relative size reflected that we expected more data and weight decay to both matter, and the lone uniform-weight arm was added just for completeness; when we saw no difference between the two weighting methods, we proceeded with the simpler uniform weights for all subsequent studies.

We ran a second study to increase data size even more, also adjusting the number of buckets into which data is subdivided. More buckets means old data is expunged more frequently, but less is expunged at a time. This lasted two weeks, with a control size of 127B words, a "500 (5)" population of 10B words, and 5B words apiece for the other three arms.

\begin{figure}[H]
    \includegraphics[width=0.5\linewidth]{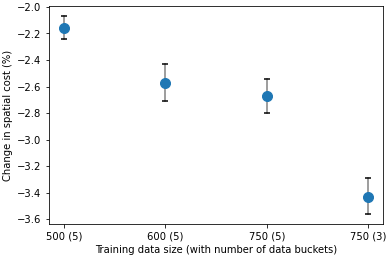}
    \caption{Training data size / buckets, round 2.}
    \label{fig:sizeandbuckets}
    \Description{Effect of training data size on spatial cost, study #2. More data gave diminishing returns in the 600-750 point range, but using coarser buckets improved cost.}
\end{figure}

In Figure \ref{fig:sizeandbuckets}, we see that more data again means a better fit, though with rapidly diminishing returns. Somewhat surprisingly, having fewer buckets led to a big drop in average spatial cost; this was a surprise, but we did not pursue it further. It is also worth noting here that these "accuracy" differences here are not large, and that they do not necessarily improve usage metrics like WPM and WMR. In fact, in this case, they did not appreciably change either. Balancing this with on-device storage costs, we somewhat arbitrarily increased the total history size and took a moderate bucket size without spending the time and resources on further exploration, settling on keeping a history of 800 taps split into 4 buckets for all subsequent studies.

\subsection{Keyboard clustering}

\begin{figure}[H]
    \includegraphics[width=0.5\linewidth]{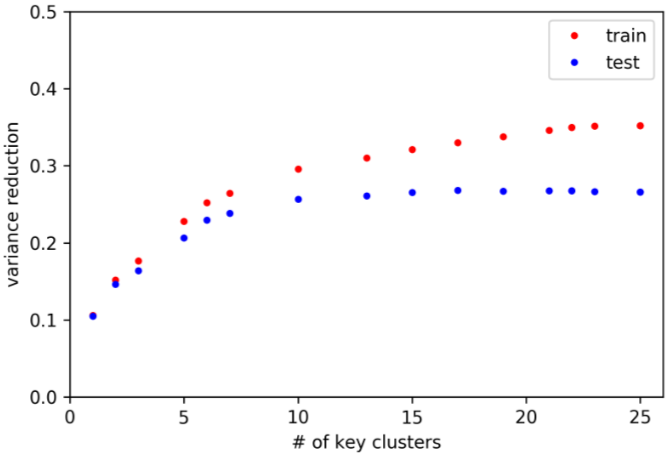}
    \caption{Cluster count vs variance reduction.}
    \label{fig:numclusters}
    \Description{When grouping all users together and computing a single fixed keyboard clustering, 7-10 clusters had the best bias-variance tradeoff on test data.}
\end{figure}

Our initial experiment with key clustering was an offline study with a very different setup that motivated the rest of this work. We had collected typing prompt data from 60 users, randomly divided the dataset into train and test sets, and greedily computed optimal global key clusters on the training data to maximize variance reduction with respect to per-cluster means. The resulting clusters and offsets were evaluated on both train and test data, finding that in that case 7 to 10 clusters was best in the bias/variance tradeoff (see Figure~\ref{fig:numclusters}).

Our live experiments trained per-user clusters rather than pooled ones, both to avoid having to send touch data off the phone and to better generalize to multiple layouts and languages. We aimed to answer whether the same number of clusters was still optimal, or whether the number mattered at all. We then ran a variety of simultaneous experiments over 18 days (control size 150B words, experiment arms 25B words each) with different numbers of target clusters to measure the tradeoff between bias and variance and to see how that affected user-visible metrics. Like the previous studies, these arms computed the appropriate per-key-cluster means and used them as offsets in the Gaussian component of the spatial model without changing isotropic scale. Since all that changes is per-cluster offsets, we may use spatial cost to compare relative accuracy of the offsets.

\begin{figure}[H]
    \includegraphics[width=0.5\linewidth]{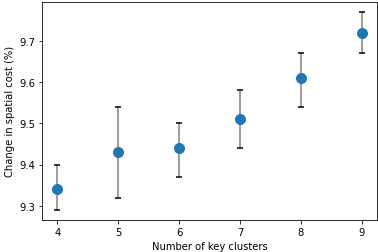}
    \caption{Cluster count vs average spatial cost.}
    \label{fig:clustersvsscore}
    \Description{Live studies with personalized keyboard clusterings showed spatial cost continuing to decrease as number of clusters decreased from 10 to 4.}
\end{figure}

The baseline experiment in Figure \ref{fig:clustersvsscore} is irrelevant here, as is the scale on the $y$-axis. What matters is that we actually see that fewer clusters lead to better fits, with the change in spatial cost often statistically significant: for 4 clusters as compared to 6 or more; for 5 , 6, or 7 clusters as compared to 8 or more; and for 8 clusters as compared to 9. But, notably, number of clusters (within this range, at least) had no statistically significant impact on the user-visible metrics of WMR or WPM, while spatial cost is not directly observable to users. From the initial offline analysis results, we had already been running our other studies with 7 clusters, so given the lack of user impact, we kept the number of clusters fixed at 7.

\subsection{Gaussian scale}

Recall the previous observation that clustering keys with similar offsets would likely result in smaller localized variance. It is therefore not enough to move offsets, but to test a variety of values of $\sigma$ for each language. All languages here had baseline values of $\sigma=0.55$ without adaptation, so we tested from $0.4$ to $0.6$ in increments of $0.05$. The control had no adaptation, and each experiment arm enabled adaptation with the previously-fixed history size of 800 taps in 4 buckets with 7 key clusters. This study ran for 26 days, and size varied based on the relative user populations for each country-language combination. The en-GB experiment, for example, had 12B words committed in the control and 1.2B per arm. The smallest population was de-DE, with 180M words per arm and 1.8B in the control.

Some example results are given below in Figures \ref{fig:gridenus}, \ref{fig:gridengb}, and \ref{fig:gridptbr}. Good values of $\sigma$ correspond to those points in Quadrant II, where WMR decreases and WPM increase. Given two results which trade off between WMR and WPM, we prefer to improve WPM; given two virtually identical results, we tend to choose the $\sigma$ value closer to the value already used in production to minimize potential disruption to other parts of the system dependent on spatial scores. 

\begin{figure}[H]
    \includegraphics[width=0.5\linewidth]{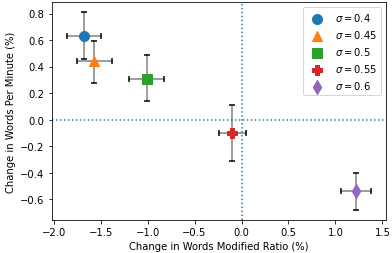}
    \caption{$\sigma$ values in {\tt en-US} outside the US.}
    \label{fig:gridenus}
    \Description{For en-US outside the US, WMR decreased and WPM increased as sigma decreased between 0.4 and 0.6.}
\end{figure}

\begin{figure}[H]
    \includegraphics[width=0.5\linewidth]{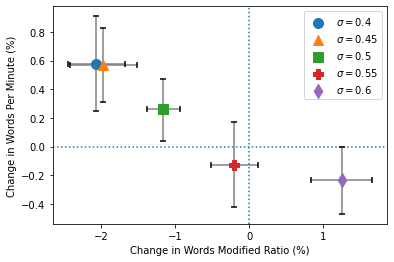}
    \caption{$\sigma$ values in {\tt en-GB}.}
    \label{fig:gridengb}
    \Description{For en-GB, WMR decreased and WPM increased as sigma decreased between 0.4 and 0.6 but with virtually no change between 0.4 and 0.45.}
\end{figure}

\begin{figure}[H]
    \includegraphics[width=0.5\linewidth]{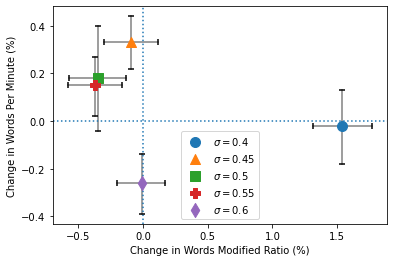}
    \caption{$\sigma$ values in {\tt pt-BR}.}
    \label{fig:gridptbr}
    \Description{For pt-BR, sigma values between 0.45 and 0.55 traded off WPM and WPR, but the extreme values of 0.4 and 0.6 both performed significantly worse.}
\end{figure}

Based on these per-language experiment results, we chose values of $\sigma$ at 0.4 for {\tt en-US} outside the US; 0.45 for {\tt en-GB}, {\tt es-ES}, {\tt es-MX}, {\tt es-US}, {\tt id-ID}, and {\tt fr-FR}; and 0.5 for {\tt en-US} in the US, {\tt de-DE}, {\tt en-IN}, {\tt it-IT}, and {\tt pt-BR}. All had previously used $\sigma=0.55$.

A perfectly fair question to ask is whether these per-language improvements come from personalization, from changing $\sigma$ (which was almost certainly suboptimal to begin with), or from a combination of the two. Changing mean offsets without changing variance typically lowers spatial costs slightly for intended decoder candidates, but may also lower costs for touches on incorrect neighboring keys; lowering variance on top of that penalizes longer distances more than before. The overall effect of the combination is to maybe increase spatial cost a bit for literal candidates, but much more for non-literal candidates.

So we ran another study with personalization disabled and a variety of $\sigma$ values, comparing to the best personalization candidate. The study ran for 13 days and collected anywhere from 90M words per experiment arm (de-DE) to 2.3B (en-US), with the control ($\sigma=0.55$) larger in every case by a factor of 10; the results are summarized in Figure \ref{tbl:compare}. Changing $\sigma$ without personalization often improved one or both of WMR and WPM as compared to the previous non-personalized baseline. But with the lone exception of English in India, the best personalization arm outperformed the best non-personalization arm in every language. That is, in each case the personalized model had a clear, statistically significant improvement in at least one of WMR and WPM over the best non-personalized model, while either improving or having no statistically significant difference in the other metric.

\begin{figure}
\caption{Metric changes (in percent) for optimal $\sigma$ values with personalization on and off.}
\label{tbl:compare}
\centering
\setlength\tabcolsep{3pt}  
\begin{tabular*}{\linewidth}{@{\extracolsep{\fill}}|c|ccccccccccc|}
\hline
Lang. & de-DE & en-GB & en-IN & en-US$^*$ & es-ES & es-MX & es-US & fr-FR & id & it-IT & pt-BR\\
\hline
$\sigma_\textrm{on}$ & 0.50 & 0.45 & 0.50 & 0.40/0.50 & 0.45 & 0.45 & 0.45 & 0.45 & 0.45 & 0.50 & 0.50\\
WMR & -1.09 & -1.88 & -0.51 & -1.63/-0.05 & -0.75 & +0.22 & -0.26 & -1.81 & -1.05 & -0.96 & -0.62 \\
WPM & +0.57 & +0.51 & +0.20 & +0.59/+0.46 & +0.63 & +0.61 & +0.43 & +0.61 & +0.42 & +0.60 & +0.28 \\
\hline
$\sigma_\textrm{off}$ & 0.50 & 0.45 & 0.50 & 0.45/0.55 & 0.45 & 0.50 & 0.50 & 0.45 & 0.50 & 0.50 & 0.50\\
WMR & -1.25 & -1.40 & -0.43 & -1.39/0 & +0.01 & +0.22 & +0.13 & -1.10 & -0.66 & -0.89 & +0.05 \\
WPM & +0.12 & +0.57 & +0.21 & +0.49/0 & +0.20 & +0.06 & +0.12 & +0.35 & +0.32 & -0.27 & +0.28 \\
\hline
\multicolumn{12}{r}{\footnotesize $^*$: en-US values are split into users geographically outside/inside the US.}
\end{tabular*}
\end{figure}

\section{Covariance}

{\it N.B.: Using full covariance did not yield any significant improvements. This section describes the attempted setup simply for the sake of completeness.}

Modeling the offset means have certainly proven useful, so it follows that we should ask about better modeling of the covariances. Our method so far has implied that, after normalizing key width and height, the distribution around a personalized key center is isotropic (i.e., symmetrically distributed radially); Figures \ref{fig:user0} and \ref{fig:user10} suggest otherwise. As one example, the touch points on the [A] key in Figure \ref{fig:user0} are skewed in a direction not parallel to either axis.

In Section \ref{sec:setup}, we defined the {\tt Stats} map and suggested that we can augment it with the per-key sums $\sum (\Delta x)^2$, $\sum (\Delta y)^2$, and $\sum (\Delta x)(\Delta y)$. We do so without changing Algorithm \ref{Algorithm} as far as cluster computation and offsets go. For a given cluster $C$, let $N = \sum_C 1$ be the number of observations in the cluster, and let
$$\vec{\mu} = (\overline{x}, \overline{y}) := (1/N)\left(\sum_C \Delta x, \sum_C \Delta y\right)$$
be the offset mean. We compute the scatter matrix with respect to $\vec{\mu}$:

\begin{align*}
S_\mu & = \frac{1}{N} \sum_C (\vec{\Delta x} - \vec{\mu}) \cdot (\vec{\Delta x} - \vec{\mu})^T\\
& = \left(\begin{array}{cc}
\sum (\Delta x)^2 - N\overline{x}^2 & \sum \Delta x \Delta y - N\overline{x}\overline{y}\\
\sum \Delta x \Delta y - N\overline{x}\overline{y} & \sum (\Delta y)^2 - N\overline{y}^2
\end{array}\right)
\end{align*}

This is the {\it maximum likelihood} estimate for the covariance of the distribution. Sparsity is an issue: if a cluster has 2 or fewer points it it is degenerate and even with 3 or more points we may have a poor estimate. Instead we use a {\it maximum a posteriori} estimate based on the conjugate prior\footnote{See e.g., \citep{murphy2012machine}, \S 4.6, equation 4.216, taking $D=2$ and $\nu_0 = D+2 = 4$.}

$$\Sigma_\mu = \frac{(1/N) diag(S_\mu) + S_\mu}{8 + N}.$$

Now we may replace the Gaussian argument of the $min$ in expression (\ref{eqn:gaussian}) using the Mahalanobis distance:

$$(D(\vec{x}))^2/2 = \frac{1}{2}(\vec{x} - \vec{\mu})^T  \cdot \Sigma_\mu^{-1} \cdot (\vec{x} - \vec{\mu})$$

To smooth between sparse and dense clusters, and to maintain control over the size of our distance estimate, we treat the entire keyboard as one "global" cluster and compute $\Sigma_{\mathrm global}$. In the event that $\Sigma_\mu$ is not invertible for a given cluster or still not trustworthy as a result of too few observations, we fall back to $\Sigma_{\mathrm global}$. In either case, we reduce $\Sigma_{\mathrm global}$ to an isotropic approximation:\footnote{This is the geometric mean of the eigenvalues of $\Sigma_{global}$.}

$$\sigma^2_{\mathrm global} := \left(\det{\Sigma_{\mathrm global}}\right)^{1/2}.$$

Given our initial isotropic parameter $\sigma_0$, we then scale the square of the Mahalanobis distance to get our new Gaussian cost $(\sigma_0^2 / \sigma_{\mathrm global}^2) D(\vec{x})^2/2$. The use of $\sigma_0$ lets us retain a tuning knob to tweak the relative impact of the Gaussian cost in the overall spatial cost, and the fallback case now behaves like the original isotropic estimate $\sigma_0^2 I$.

Using this formulation, we ran a new study. The control used personalized key centers but isotropic covariance $\sigma_0^2 I$, while each experiment added personalized covariance with a different $\sigma_0$ base value. The most promising arms for each language were then extracted to a second study which ran longer and on a larger fraction of users to get more reliable results. Unfortunately, only {\tt en-GB} with $\sigma_0=0.45$ showed a statistically-significant improvement in WMR, but only just barely ($-0.10\%$). No languages showed a statistically-significant improvement in WPM, and a few were slightly worse. Since this would be minimal gain at best for the added complexity, we did not proceed further with personalized covariance in any language.

\section{Limitations and future work}

Our most important finding here is a definitive improvement in typing speed and error rate across all major languages studied when using personalized key centers for touch typing. We have confidence in the results because they are comprehensive, measuring metric changes across millions or billions of words typed by uniform random samples of users in their real, representative keyboard usage. Due to the size of the parameter space, we may not have a globally-optimal model, but we have nonetheless seen statistically significant improvement from our locally-optimized one.

The magnitude of these changes is perhaps smaller than in previous related work, such as \citep{weir2012user}, because of imposed product requirements. For example, Gboard may have more complex language models and competing systems like auto-correction modules that work to reduce mistakes before the introduction of a personalized spatial model. Although we might consider measuring the impact on a more level playing field, we cannot, for example, disable the auto-correction module in both control and experiment populations because the greatly-reduced typing quality would be considered harmful to our users. Another such limitation is the bucketing scheme used to prevent possible reconstruction of words typed: user privacy is a hard requirement that we cannot disable, even if it might simplify our code and give us better control over the history.

The most important direction for future work, following \citep{goel2013contexttype} and \citep{music2016nomadic}, would be incorporation of posture and orientation into the model. We never ask users for their posture or have a reliable way to infer it, so our setup here was simplified by not relying on posture-predicting models whose accuracy we would be unable to evaluate. But incorporating orientation would be a simple next step, computing separate statistics maps for portrait and landscape data and even slicing typing metrics by orientation; work to detect posture and divert accordingly could follow. The prior literature suggests that per-posture personalized models could be a sizable improvement over posture-agnostic ones.

Another likely direction for future work here is what we specifically avoided discussing until now: gesture decoding most likely has ample room for improvements due to personalization, but is a very different style of input with a different spatial model format, and it will likely require a materially different form of adaptation. For example, we have seen that touch typing tends to follow a Gaussian distribution per key, but gesture typing is likely to have error dependent on the location of previous and next keys. (See also \citep{quinn2018modeling} for discussion of gesture glide paths, especially Figure 6 for a variety of possible corner shapes that do not apply to touch typing.)

\section{Conclusion}

Gboard users do not generally type by pressing directly on physical centers of keys. A Gaussian spatial model which accounted for personalized typing behavior showed modest --- but real --- improvements in both error rates and typing speed, largely by reducing both false positives and false negatives in auto-correction decisions.

We found that more training data generally leads to better fits for user touch behavior, though decaying weights seem to make no difference, and quickly diminishing returns mean that better fits do not necessarily lead to improved typing quality. We also found that the number of key clusters used to avoid overfitting very slightly affects model accuracy, but that any number of clusters within the range tested did not produce measurable user-visible changes.

We did not capture the effect of changes in typing posture or device orientation. If we had seen better performance with less training data or with training weight decay, we might have concluded that we were effectively adjusting to these changes quickly, but this was not the case. Future work should include separate modeling for these, but we did show that a single per-user model could give significant improvements without even considering them.

Last, we found that using personalized covariance matrices might better model user typing habits, but did not improve typing accuracy or speed beyond the gains from personalized offset means alone.

\begin{acks}
We would like to thank Petr Zadrazil for work on the adaptation framework used here, Billy Dou and Leif Johnson for preliminary analysis and earlier efforts on the spatial model personalization problem, and Vlad Schogol for a great many discussions and suggestions along the way.
\end{acks}

\appendix

\section{Clustering algorithm}

See \citep{breiman1984classification} for general discussion of decision trees and Algorithm \ref{Algorithm}\label{app:algorithm} for our specific clustering pseudocode.

\begin{algorithm*}
\caption{Computes clusters of keys and personalized offsets for each cluster.} 
\label{Algorithm}

\begin{algorithmic}
\STATE Initialize $tree$ to a single root node representing the cluster containing all keys.
\WHILE{$|{\tt leaves}(tree)| < K$ and some cluster has at least 2 points}
    \STATE $bestReduction \mapsfrom 0$, $bestSplit \mapsfrom None$, $bestLeaf \mapsfrom None$
    \FORALL{leaf, cluster $\in {\tt leaves}(tree)$}
        \FORALL{$split \in \mathtt{horizSplits}(\mathrm{cluster}) \ \bigcup\  \mathtt{vertSplits}(\mathrm{cluster})$}
            \STATE Let $C_*$, $C_1$, $C_2$ represent cluster and its two subclusters as separated by $split$.
            \STATE Let $reduction := f(C_1) + f(C_2) - f(C_*)$, where $f(C) := |C| \cdot (E[\Delta x]^2 + E[\Delta y]^2)$
            \IF{$reduction > bestReduction$}
              \STATE $bestReduction \mapsfrom reduction$
              \STATE $bestSplit \mapsfrom split$
              \STATE $bestLeaf \mapsfrom leaf$
            \ENDIF
        \ENDFOR
    \ENDFOR
    \STATE Split the cluster at $bestLeaf$ into two child clusters using $bestSplit$
\ENDWHILE
\FORALL{cluster $\in {\tt leaves}(tree)$}
    \STATE The offset for keys in this cluster is $(E[\Delta x], E[\Delta y])$.
\ENDFOR
\end{algorithmic}
\end{algorithm*}

The algorithm consists of a sequence of greedy partitions along vertical or horizontal lines using mean square error. The key piece to derive here is the line:
$$reduction = f(C_1) + f(C_2) - f(C_*)$$
where $C_*$ is the disjoint union of $C_1$ and $C_2$ and 
$$f(C) = |C| \cdot(E[\Delta x]^2 + E[\Delta y]^2).$$
Since this uses square error, we can split the sums into one sum which is a function of $x$-coordinates and one which is a function of $y$-coordinates. Write $z$ for either the $\Delta x$ terms or the $\Delta y$ terms. For a cluster $C = \{z_1, \ldots, z_n\}$, the value of $s$ minimizing
$$V_C(s) := \frac{1}{n} \sum_{i=1}^n (z_i - s)^2$$
is well-known and easily shown to be $s = E[z]$, with MSE given by
$$V_C(E[z]) = Var(z) = E[z^2] - E[z]^2.$$
Scaling by $|C|$, we can write the first term as:
\begin{align*}
|C_*| \cdot E_*[z^2] & = \sum_{C_*} z^2\\
& = \sum_{C_1} z^2 + \sum_{C_2} z^2\\
& = |C_1| \cdot E_1 [z^2] + |C_2| \cdot E_2[z^2].
\end{align*}
These terms then cancel out in $reduction$, and we're left with:
$$|C_*| V_{C_*} - |C_1| V_{C_1} - |C_2| V_{C_2} = |C_1| E_1[z]^2 + |C_2| E_2[z^2] - |C_*| E_*[z^2].$$

Substituting $\Delta x$ and $\Delta y$ in turn for $z$ and adding gives the form of $reduction$ in the algorithm.

It's worth noting that this algorithm is efficient: within each cluster, we keep a sorted list of a horizontal split points and partial sums of the {\tt Stats} map, and similarly for the vertical split points. Then iterating through the $m$ split points in order takes $O(m)$ time, and splitting a cluster after finding the best point takes $O(m)$ time for the same reason. In each iteration of the greedy algorithm, we need only recompute reduction values for newly-split clusters, and only in the direction orthogonal to the split. It follows that for a keyboard layout with $N$ keys and $K$ clusters, we can produce the decision tree in $O(N \log N + N K)$ time.

\bibliographystyle{ACM-Reference-Format}
\bibliography{references}


\begin{thebibliography}{19}


\ifx \showCODEN    \undefined \def \showCODEN     #1{\unskip}     \fi
\ifx \showDOI      \undefined \def \showDOI       #1{#1}\fi
\ifx \showISBNx    \undefined \def \showISBNx     #1{\unskip}     \fi
\ifx \showISBNxiii \undefined \def \showISBNxiii  #1{\unskip}     \fi
\ifx \showISSN     \undefined \def \showISSN      #1{\unskip}     \fi
\ifx \showLCCN     \undefined \def \showLCCN      #1{\unskip}     \fi
\ifx \shownote     \undefined \def \shownote      #1{#1}          \fi
\ifx \showarticletitle \undefined \def \showarticletitle #1{#1}   \fi
\ifx \showURL      \undefined \def \showURL       {\relax}        \fi
\providecommand\bibfield[2]{#2}
\providecommand\bibinfo[2]{#2}
\providecommand\natexlab[1]{#1}
\providecommand\showeprint[2][]{arXiv:#2}

\bibitem[\protect\citeauthoryear{Azenkot and Zhai}{Azenkot and Zhai}{2012}]%
        {azenkot2012touch}
\bibfield{author}{\bibinfo{person}{Shiri Azenkot} {and} \bibinfo{person}{Shumin
  Zhai}.} \bibinfo{year}{2012}\natexlab{}.
\newblock \showarticletitle{Touch behavior with different postures on soft
  smartphone keyboards}. In \bibinfo{booktitle}{\emph{Proceedings of the 14th
  international conference on Human-computer interaction with mobile devices
  and services}}. \bibinfo{pages}{251--260}.
\newblock


\bibitem[\protect\citeauthoryear{Baldwin and Chai}{Baldwin and Chai}{2012}]%
        {baldwin2012towards}
\bibfield{author}{\bibinfo{person}{Tyler Baldwin} {and} \bibinfo{person}{Joyce
  Chai}.} \bibinfo{year}{2012}\natexlab{}.
\newblock \showarticletitle{Towards online adaptation and personalization of
  key-target resizing for mobile devices}. In
  \bibinfo{booktitle}{\emph{Proceedings of the 2012 ACM international
  conference on Intelligent User Interfaces}}. \bibinfo{pages}{11--20}.
\newblock


\bibitem[\protect\citeauthoryear{Breiman, Friedman, Stone, and Olshen}{Breiman
  et~al\mbox{.}}{1984}]%
        {breiman1984classification}
\bibfield{author}{\bibinfo{person}{Leo Breiman}, \bibinfo{person}{Jerome
  Friedman}, \bibinfo{person}{Charles~J Stone}, {and}
  \bibinfo{person}{Richard~A Olshen}.} \bibinfo{year}{1984}\natexlab{}.
\newblock \bibinfo{booktitle}{\emph{Classification and regression trees}}.
\newblock \bibinfo{publisher}{CRC press}.
\newblock


\bibitem[\protect\citeauthoryear{Buschek, Bisinger, and Alt}{Buschek
  et~al\mbox{.}}{2018}]%
        {buschek2018researchime}
\bibfield{author}{\bibinfo{person}{Daniel Buschek}, \bibinfo{person}{Benjamin
  Bisinger}, {and} \bibinfo{person}{Florian Alt}.}
  \bibinfo{year}{2018}\natexlab{}.
\newblock \showarticletitle{ResearchIME: A mobile keyboard application for
  studying free typing behaviour in the wild}. In
  \bibinfo{booktitle}{\emph{Proceedings of the 2018 CHI Conference on Human
  Factors in Computing Systems}}. \bibinfo{pages}{1--14}.
\newblock


\bibitem[\protect\citeauthoryear{Dhakal, Feit, Kristensson, and
  Oulasvirta}{Dhakal et~al\mbox{.}}{2018}]%
        {dhakal2018observations}
\bibfield{author}{\bibinfo{person}{Vivek Dhakal}, \bibinfo{person}{Anna~Maria
  Feit}, \bibinfo{person}{Per~Ola Kristensson}, {and} \bibinfo{person}{Antti
  Oulasvirta}.} \bibinfo{year}{2018}\natexlab{}.
\newblock \showarticletitle{Observations on typing from 136 million
  keystrokes}. In \bibinfo{booktitle}{\emph{Proceedings of the 2018 CHI
  Conference on Human Factors in Computing Systems}}. \bibinfo{pages}{1--12}.
\newblock


\bibitem[\protect\citeauthoryear{Goel, Findlater, and Wobbrock}{Goel
  et~al\mbox{.}}{2012}]%
        {goel2012walktype}
\bibfield{author}{\bibinfo{person}{Mayank Goel}, \bibinfo{person}{Leah
  Findlater}, {and} \bibinfo{person}{Jacob Wobbrock}.}
  \bibinfo{year}{2012}\natexlab{}.
\newblock \showarticletitle{WalkType: using accelerometer data to accomodate
  situational impairments in mobile touch screen text entry}. In
  \bibinfo{booktitle}{\emph{Proceedings of the SIGCHI Conference on Human
  Factors in Computing Systems}}. \bibinfo{pages}{2687--2696}.
\newblock


\bibitem[\protect\citeauthoryear{Goel, Jansen, Mandel, Patel, and
  Wobbrock}{Goel et~al\mbox{.}}{2013}]%
        {goel2013contexttype}
\bibfield{author}{\bibinfo{person}{Mayank Goel}, \bibinfo{person}{Alex Jansen},
  \bibinfo{person}{Travis Mandel}, \bibinfo{person}{Shwetak~N Patel}, {and}
  \bibinfo{person}{Jacob~O Wobbrock}.} \bibinfo{year}{2013}\natexlab{}.
\newblock \showarticletitle{ContextType: using hand posture information to
  improve mobile touch screen text entry}. In
  \bibinfo{booktitle}{\emph{Proceedings of the SIGCHI conference on human
  factors in computing systems}}. \bibinfo{pages}{2795--2798}.
\newblock


\bibitem[\protect\citeauthoryear{Henze, Rukzio, and Boll}{Henze
  et~al\mbox{.}}{2012}]%
        {henze2012observational}
\bibfield{author}{\bibinfo{person}{Niels Henze}, \bibinfo{person}{Enrico
  Rukzio}, {and} \bibinfo{person}{Susanne Boll}.}
  \bibinfo{year}{2012}\natexlab{}.
\newblock \showarticletitle{Observational and experimental investigation of
  typing behaviour using virtual keyboards for mobile devices}. In
  \bibinfo{booktitle}{\emph{Proceedings of the SIGCHI Conference on Human
  Factors in Computing Systems}}. \bibinfo{pages}{2659--2668}.
\newblock


\bibitem[\protect\citeauthoryear{Jelinek}{Jelinek}{1998}]%
        {jelinek1998statistical}
\bibfield{author}{\bibinfo{person}{Frederick Jelinek}.}
  \bibinfo{year}{1998}\natexlab{}.
\newblock \bibinfo{booktitle}{\emph{Statistical methods for speech
  recognition}}.
\newblock \bibinfo{publisher}{MIT press}.
\newblock


\bibitem[\protect\citeauthoryear{Murphy}{Murphy}{2012}]%
        {murphy2012machine}
\bibfield{author}{\bibinfo{person}{Kevin~P Murphy}.}
  \bibinfo{year}{2012}\natexlab{}.
\newblock \bibinfo{booktitle}{\emph{Machine learning: a probabilistic
  perspective}}.
\newblock \bibinfo{publisher}{MIT press}.
\newblock


\bibitem[\protect\citeauthoryear{Musi{\'c} and Murray-Smith}{Musi{\'c} and
  Murray-Smith}{2016}]%
        {music2016nomadic}
\bibfield{author}{\bibinfo{person}{Josip Musi{\'c}} {and}
  \bibinfo{person}{Roderick Murray-Smith}.} \bibinfo{year}{2016}\natexlab{}.
\newblock \showarticletitle{Nomadic input on mobile devices: the influence of
  touch input technique and walking speed on performance and offset modeling}.
\newblock \bibinfo{journal}{\emph{Human--Computer Interaction}}
  \bibinfo{volume}{31}, \bibinfo{number}{5} (\bibinfo{year}{2016}),
  \bibinfo{pages}{420--471}.
\newblock


\bibitem[\protect\citeauthoryear{Ouyang, Rybach, Beaufays, and Riley}{Ouyang
  et~al\mbox{.}}{2017}]%
        {ouyang2017mobile}
\bibfield{author}{\bibinfo{person}{Tom Ouyang}, \bibinfo{person}{David Rybach},
  \bibinfo{person}{Fran{\c{c}}oise Beaufays}, {and} \bibinfo{person}{Michael
  Riley}.} \bibinfo{year}{2017}\natexlab{}.
\newblock \showarticletitle{Mobile keyboard input decoding with finite-state
  transducers}.
\newblock \bibinfo{journal}{\emph{arXiv preprint arXiv:1704.03987}}
  (\bibinfo{year}{2017}).
\newblock


\bibitem[\protect\citeauthoryear{Palin, Feit, Kim, Kristensson, and
  Oulasvirta}{Palin et~al\mbox{.}}{2019}]%
        {palin2019people}
\bibfield{author}{\bibinfo{person}{Kseniia Palin}, \bibinfo{person}{Anna~Maria
  Feit}, \bibinfo{person}{Sunjun Kim}, \bibinfo{person}{Per~Ola Kristensson},
  {and} \bibinfo{person}{Antti Oulasvirta}.} \bibinfo{year}{2019}\natexlab{}.
\newblock \showarticletitle{How do people type on mobile devices? Observations
  from a study with 37,000 volunteers}. In
  \bibinfo{booktitle}{\emph{Proceedings of the 21st International Conference on
  Human-Computer Interaction with Mobile Devices and Services}}.
  \bibinfo{pages}{1--12}.
\newblock


\bibitem[\protect\citeauthoryear{Quinn and Zhai}{Quinn and Zhai}{2018}]%
        {quinn2018modeling}
\bibfield{author}{\bibinfo{person}{Philip Quinn} {and} \bibinfo{person}{Shumin
  Zhai}.} \bibinfo{year}{2018}\natexlab{}.
\newblock \showarticletitle{Modeling gesture-typing movements}.
\newblock \bibinfo{journal}{\emph{Human--Computer Interaction}}
  \bibinfo{volume}{33}, \bibinfo{number}{3} (\bibinfo{year}{2018}),
  \bibinfo{pages}{234--280}.
\newblock


\bibitem[\protect\citeauthoryear{Reyal, Zhai, and Kristensson}{Reyal
  et~al\mbox{.}}{2015}]%
        {reyal2015performance}
\bibfield{author}{\bibinfo{person}{Shyam Reyal}, \bibinfo{person}{Shumin Zhai},
  {and} \bibinfo{person}{Per~Ola Kristensson}.}
  \bibinfo{year}{2015}\natexlab{}.
\newblock \showarticletitle{Performance and user experience of touchscreen and
  gesture keyboards in a lab setting and in the wild}. In
  \bibinfo{booktitle}{\emph{Proceedings of the 33rd Annual ACM Conference on
  Human Factors in Computing Systems}}. \bibinfo{pages}{679--688}.
\newblock


\bibitem[\protect\citeauthoryear{Tang, Agarwal, O'Brien, and Meyer}{Tang
  et~al\mbox{.}}{2010}]%
        {tang2010overlapping}
\bibfield{author}{\bibinfo{person}{Diane Tang}, \bibinfo{person}{Ashish
  Agarwal}, \bibinfo{person}{Deirdre O'Brien}, {and} \bibinfo{person}{Mike
  Meyer}.} \bibinfo{year}{2010}\natexlab{}.
\newblock \showarticletitle{Overlapping experiment infrastructure: More,
  better, faster experimentation}. In \bibinfo{booktitle}{\emph{Proceedings of
  the 16th ACM SIGKDD international conference on Knowledge discovery and data
  mining}}. \bibinfo{pages}{17--26}.
\newblock


\bibitem[\protect\citeauthoryear{Weir, Rogers, Murray-Smith, and
  L{\"o}chtefeld}{Weir et~al\mbox{.}}{2012}]%
        {weir2012user}
\bibfield{author}{\bibinfo{person}{Daryl Weir}, \bibinfo{person}{Simon Rogers},
  \bibinfo{person}{Roderick Murray-Smith}, {and} \bibinfo{person}{Markus
  L{\"o}chtefeld}.} \bibinfo{year}{2012}\natexlab{}.
\newblock \showarticletitle{A user-specific machine learning approach for
  improving touch accuracy on mobile devices}. In
  \bibinfo{booktitle}{\emph{Proceedings of the 25th annual ACM symposium on
  User interface software and technology}}. \bibinfo{pages}{465--476}.
\newblock


\bibitem[\protect\citeauthoryear{Zhai and Kristensson}{Zhai and
  Kristensson}{2012}]%
        {zhai2012word}
\bibfield{author}{\bibinfo{person}{Shumin Zhai} {and} \bibinfo{person}{Per~Ola
  Kristensson}.} \bibinfo{year}{2012}\natexlab{}.
\newblock \showarticletitle{The word-gesture keyboard: reimagining keyboard
  interaction}.
\newblock \bibinfo{journal}{\emph{Commun. ACM}} \bibinfo{volume}{55},
  \bibinfo{number}{9} (\bibinfo{year}{2012}), \bibinfo{pages}{91--101}.
\newblock


\bibitem[\protect\citeauthoryear{Zhu, Luo, Bi, and Zhai}{Zhu
  et~al\mbox{.}}{2018}]%
        {zhu2018typing}
\bibfield{author}{\bibinfo{person}{Suwen Zhu}, \bibinfo{person}{Tianyao Luo},
  \bibinfo{person}{Xiaojun Bi}, {and} \bibinfo{person}{Shumin Zhai}.}
  \bibinfo{year}{2018}\natexlab{}.
\newblock \showarticletitle{Typing on an invisible keyboard}. In
  \bibinfo{booktitle}{\emph{Proceedings of the 2018 CHI Conference on Human
  Factors in Computing Systems}}. \bibinfo{pages}{1--13}.
\newblock


\end{thebibliography}

\received{February 2022}
\received[revised]{May 2022}
\received[accepted]{June 2022}

\end{document}